# A Probabilistic Simulation Based VaR Computation and Sensitivity Analysis Method

Wendy Li

## ABSTRACT

This paper presents a new method to compute VaR (value at risk) and perform corresponding variance based sensitivity analysis. VaR has a long history of being applied in stock price prediction and investment portfolio analysis. Traditional method, however, is mainly analytical, and has certain limitations. The VaR simulation, on the other hand, provides more realistic analysis but is very slow which affects the applications. This study proposes a new VaR computation method based on a probabilistic simulation technique called Simulation As You Operate (SAYO). It is always helpful to know the most influential factors in an investment, and thus a sensitivity analysis method based on SAYO is also introduced to enhance investment analysis.

## INTRODUCTION

VaR (Value at Risk) is widely used to investigate the risk (especially risk of loss) on an investment portfolio with one or more financial assets over the given time horizon (Jorion 1997). Traditional calculation of VaR is analytical based, especially using variance-covariance method (Linsmeier and Pearson 2000). But analytical method has certain drawbacks. First, Analytical VaR assumes not only that the historical returns follow a normal distribution, but also that the changes in price of the assets included in the portfolio follow a normal distribution. And this very rarely survives the test of reality (Rockafellar and Uryasev 2002). Second, Analytical VaR does not cope very well with securities that have a non-linear payoff distribution like options or mortgage-backed securities (Pflug 2000). Finally, if our historical series exhibits heavy tails, then computing Analytical VaR using a normal distribution will underestimate VaR at high confidence levels and overestimate VaR at low confidence levels (Duffie and Pan 1997). As an alternative to analytical VaR, Monte Carlo simulation is used. But it has been noticed that conventional Monte Carlo simulation becomes unbearably slow when the number of variables increase exponentially (Du et al. 2014). This study therefore proposes a new probabilistic based method to improve the user experience of performing VaR analysis using Real-time Monte Carlo simulation, and to realize faster VaR calculation.

**LITERATURE REVIEW**

**Simulation in Decision Making**

Many of the features, events and processes which control the behavior of currently available complex systems will not be known or understood with certainty (Du and El-Gafy 2010). This is because, for most real-world systems, at least some of the controlling parameters, processes and events are often stochastic, uncertain and/or poorly understood (Du and El-Gafy 2011). The objective of many decision support systems is to identify and quantify the risks associated with a particular option, plan or design. Incorporating uncertainties into the analysis of system behavior is called uncertainty analysis. Uncertainty analysis is part of every decision we make. We are constantly faced with uncertainty, ambiguity, and variability. And even though we have unprecedented access to information, we can't accurately predict the future. Simulation, in this case, is a possible solution which lets us visualize all the possible outcomes of the decisions and assess the impact of risk, allowing for better decision making under uncertainty. Simulating a system in the face of such uncertainty and quantifying such risks requires that the uncertainties be quantitatively included in the calculations.

The simulation method has been widely used to tackle problems in multiple areas including engineering (Du and El-Gafy 2012; Du and El-Gafy 2014; Du and El-Gafy 2014; Du et al.), management (Du et al. 2012), planning (Du and Liu), information technology (Du et al. 2014), financing (Du 2014; Du and Bormann 2014; Glasserman 2003), healthcare (Hay et al. 1987), serious gaming (Liu et al. 2014; Liu et al. 2014) and urban planning (Du and Wang 2011). This study utilizes a latest probabilistic method that enables real-time Monte Carlo simulation.

**Real-time Monte Carlo Simulation**

This paper is based on a method called "Simulate As You Operate" (SAYO) developed by Du and colleagues (Du et al. 2014). The following describes their method in brief. A model is a reproduction of a real world problem *P*. Under SAYO, a model can be defined as a collection of *M* random variables (M>=2), and the operations over them, denoted as *F*, including arithmetical operations, logic operations, matrix operations and so on. The result of the model simulation is denoted as *R*. Therefore:

$$R = F(P) \qquad (1)$$

The original problem is either aggregatable or nested. Aggregatable means the original problem space can be projected onto at least two sub-spaces which are independent of each other. From the perspective of practical application, aggregatable problems have at least two independent parts (sub-problems) such that each part can be simulated independently and in parallel, and the results can be synthesized later. Suppose a problem *P* has *M* variables:

$$P = \{x_1, x_2, \dots, x_M\} \qquad (2)$$

In other words, *P* belongs to an M-dimensional space:

$$P \in \mathfrak{R}^M \tag{3}$$

$P$ is divisible if it can be projected into $k$ sub-problems, and each sub-problem is embedded in a $m_i$ dimensional space, i.e.,

$$P = \{p_1, p_2, \ldots p_k\}, \text{for any } i \text{ and } j \leq k, p_i \notin p_j \text{ and } p_j \notin p_i \tag{4}$$

$$p_i \in \mathfrak{R}^{m_i} \tag{5}$$

$$M = \sum_{i=1}^{k} m_i \tag{6}$$

The above situation is called incompletely divisible. When the $m_1$ variables of sub-problem 1 have been parameterized, a probabilistic simulation may be executed immediately. Denote $f_i$ as the conversion function that yields the simulation result $r_i$ of sub-problem $p_i$, then the above process can be described as:

$$r_1 = f_1(p_1) \text{ at } sim_1 \tag{7}$$

Observe that sim1 occurs when the user is still parameterizing $p_2$. The parameterization process is executed in parallel with the random number generation (RNG) processes and simulation processes. This process will continue until the entire problem or divisible part of the problem is simulated. Then the simulation result $R$ of the complete problem $P$ can be written as:

$$R = \{r_1, r_2, \ldots r_k\}, \text{ where}$$

$$r_1 = f_1(p_1) \text{ at } sim_1$$

$$r_2 = f_2(p_2) \text{ at } sim_2 \tag{8}$$

$$\ldots\ldots$$

$$r_k = f_k(p_k) \ at \ sim_k$$

instead of

$$R = F(P) \ at \ sim_{all}, \ sim_{all} = \sum_{i=1}^{k} sim_i \qquad (9)$$

An extreme case of the divisible problem would be projecting the original problem $P$ onto $K$ sub-spaces, where each sub-space only has 1 dimension, or:

$$p_i \in \Re^1 \qquad (10)$$

Therefore:

$$K = M = \sum_{i=1}^{k} mi \qquad (11)$$

This situation is called completely divisible. In this situation, each variable of the problem will be ready for simulation after the user parameterized and defined part of the model. The basic unit of simulation occurs between two variables.

In another situation, the problem has a nested structure. Suppose the problem $P$ can be projected into a set of sub-spaces:

$$P = \{P_\alpha, P_\beta, P_\gamma, \ldots, P_\delta\} \qquad (12)$$

For $p+q+l+\ldots+n=M$, $p>=0$, $q>=0$, $l>=0,\ldots,n>=0$:

$$P_\alpha = \{x_{\alpha 1}, x_{\alpha 2}, \dots, x_{\alpha p}\}$$

$$P_\beta = \{x_{\beta 1}, x_{\beta 2}, \dots, x_{\beta q}\}$$

$$P_\gamma = \{x_{\gamma 1}, x_{\gamma 2}, \dots, x_{\gamma l}\}$$

$$\dots$$

$$P_\delta = \{x_{\delta 1}, x_{\delta 2}, \dots, x_{\delta n}\} \qquad (13)$$

For each $x_{\alpha i} \in P_\alpha$,

$$x_{\alpha i} = \{x_{\beta 1}, x_{\beta 2}, \dots, x_{\beta k}\}, k \leq q \qquad (14)$$

And for each $x_{\beta j} \in P_\beta$,

$$x_{\beta j} = \{x_{\gamma 1}, x_{\gamma 2}, \dots, x_{\gamma g}\}, g \leq l \qquad (15)$$

Until $P_\delta$ has been defined. In the nesting case, $P_\delta$ will first be defined and parameterized, and then the lower level relative of $P_\delta$ will be simulated based on the outcomes of $P_\delta$. This process will be repeated in parallel with the model definition and parameterization process without any interruptions until the bottom level $P_\alpha$ has been defined, parameterized and simulated. The real-time simulation for nesting problems is realized.

The total probabilistic simulation time is divided into four components:

- **Parameterization time (PT or pt$_i$):** The time spent by the user to parameterize the model. For example, the user defines the probability density functions (PDFs) of the model inputs;

- **Random number generation time (GT or $gt_i$):** The time spent by the system to generate or retrieve random numbers for the simulation according to the arbitrary distributions defined by the user;

- **Simulation time (ST or $st_i$):** The pure time spent by the system to perform the actual simulation tasks; and

- **Overhead (OH or $ot_i$):** Simulation involves lots of data fetching and processing operations and transferring. The data needs to be read and saved in the computer memory hierarchy frequently. Typically, a lot of time is required to transfer data between central processing unit (CPU) and the main memory, between CPU and secondary storage (hard disk), off-line storage and tertiary storage (e.g. tape drives), and among different hierarchical levels of the memory system. From the database operation standpoint, time is also required to perform database operations such as database initialization, read/write, insert, update, delete, merge and indexing etc. The time consumed in such operations does not directly contribute to the probabilistic simulation, and thus can be called overhead.

In SATO, the four components of a probabilistic simulation can be executed in parallel to improve the efficiency. However, the extent to which the four components can be concurrently executed varies for different types of problem. For **completely divisible** problems the parameterization, generation, sub-simulation (and its corresponding overhead) can be executed concurrently. For each sub-simulation then, the time depends on the maximum of the above three. Assuming the parameterization takes the longest time then total time required for the simulation of a completely divisible problem is:

$$TT_{cd} = \sum_{i=1}^{m} \max\{pt_i, gt_i, (st_{i+1} + ot_{i+1})\} = \sum_{i=1}^{m} pt_i = PT \qquad (16)$$

where *m* equals to the number of model variables and *PT* denotes the total time for parameterization. For an **incompletely divisible problem**, the time required for the synthesis simulation, denoted as *STex* (external simulation), and its corresponding overhead, denoted as *OHex* (external overhead), need to be included to obtain the total time required for the simulation of an incompletely divisible problem:

$$TT_{icd} = \sum_{j=1}^{k} \max\{\sum_{i=1}^{m_j} pt_{ji}, \sum_{i=1}^{m_j} gt_{ji}, (st_{j-1} + ot_{j-1})\} + st_j + ot_j + ST_{ex} + OH_{ex}$$

$$= \sum_{j=1}^{k}\sum_{i=1}^{m_j} pt_{ji} + st_j + ot_j + ST_{ex} + OH_{ex} = PT + ST_{ex} + OH_{ex} + st_k + ot_k$$

$$(17)$$

As above, the significant reduction in simulation time makes SAYO real time.

**DESCRIPTION OF THE METHOD**

Based on Du's SAYO method, I propose a method for VaR computation called SAYO-VaR. Suppose an investor wants to study the investment portfolio with N stocks. SAYO-VaR maintains the PDFs of all commonly used stocks.

| $ 10,308.01 | $ 22,482.30 | $ 4,816.38 | $ (94,606.42) |
| $ (7,578.04) | $ 124,868.18 | $ 85,410.57 | $ (65,800.81) |
| $ (18,766.73) | $ 45,236.88 | $ 65,146.35 | $ 150,891.87 |
| $ (51,790.62) | $ (33,162.38) | $ (11,170.05) | $ 144,440.09 |
| $ (11,868.53) | $ (22,374.19) | $ (31,009.11) | $ 209,901.37 |
| $ (12,528.04) | $ 19,993.89 | $ (21,314.44) | $ 257,481.19 |
| $(125,012.56) | $ (34,468.30) | $ 31,582.07 | $ 148,281.07 |
| … | … | … | … |
| $ 43,787.36 | $ 20,429.10 | $ 22,226.99 | $ 5,050.66 |
| $ (12,567.33) | $ 91,789.05 | $ (36,954.05) | $ 159,183.53 |
| $ 30,083.22 | $ 53,680.95 | $ 76,680.99 | $ 232,858.93 |
| $ 12,637.63 | $ 57,313.79 | $ 66,292.69 | $ 42,832.02 |
| $ 13,721.07 | $ (4,900.02) | $ 7,784.39 | $ 96,619.77 |
| $ 93,842.83 | $ (26,135.69) | $ 79,366.73 | $ 127,081.51 |
| $ 55,239.39 | $ (77,256.76) | $ 62,160.07 | $ 131,045.87 |

**Fig. 1** SAYO-VaR operations

Referring to **Fig.1**, once the user selects a stock, the PDF pertaining to that stock will be retrieved from the database and a random number tuple will be generated to represent all possible returns by investing this stock. The VaR for the stock, as well as other alternatives such as CVaR (Conditional Value at Risk) and EVaR (Entropic Value at Risk), will be calculated and displayed instantaneously under provided time horizon and significance level α. Then the user starts to select the second stock. Similarly, a random number tuple will be generated according to the retrieved PDF to represent possible returns by investing the second stock, and the VaR, CVaR and EVaR for the second stock will be calculated and displayed instantaneously under provided time horizon and significance level α. If there are correlations among stocks, methods for preserving the correlations will be used such as Cholsky decomposition. Moreover, given the relative share of the first stock and the second stock which are provided by the user, an additive operation will be performed between random number tuples of the first and the second stocks' returns. The VaR, CVaR and EVaR of the aggregated random number tuples will be calculated and displayed instantaneously to represent portfolio risk. This process will be repeated for the rest stock selections and the VaR, CVaR and EVaR of the portfolio return will be calculated concurrently and updated on a timely basis, i.e., every time when any part of the portfolio is updated. Once the user finishes the selection of the last stock and the parameterization of the stock share on the portfolio, The VaR, CVaR and EVaR of the entire portfolio will be displayed instantaneously. SAYO is realized.

In another case, the random number tuples may be obtained directly from the stock transaction history. SAYO-VaR acquires stock transaction data from data providers and saves it on the database server. User selects certain stocks and defines the time horizon that is of the interest, for example, transactions of every minute in past 6 months, and then SAYO-VaR will retrieve relevant transaction data from the database and saves it on the temporary storage or cache. The VaR, CVaR and EVaR of each stock and the portfolio will be calculated and updated in a SAYO fashion.

In another case, the user may want to perform sensitivity analysis to check what stocks are more influential to the portfolio's final return. Instead of the traditional sensitivity analysis method where random numbers are generated for each trial for each model input, and output results are aggregated finally to calculate the sensitivity indices of each input, SAYO-VaR adopts the SAYO strategy. For example, in order to calculate Sobol's total sensitivity indices (TSI), the variance of inputs and outputs need to be calculated repeatedly on a timely fashion. BGBC enables a faster implementation of Sobol's TSI calculation. Sobol's TSI method assumes a nonlinear function can be decomposed to summands of orthogonal increasing order terms which is called ANOVA-representation:

$$f(x_1, x_2, ..., x_m) = f_0 + \sum_{i=1}^{m} f_i(x_i) + \sum_{i_1=1}^{m} \sum_{i_2=i_1+1}^{m} f_{i_1 i_2}(x_{i_1}, x_{i_2}) + ... + f_{1...m}(x_1, ..., x_m)$$
(18)

Assume $x_i (i=1,2,...m)$ are independent random variables with probability density functions $p_i(x_i)$, then the constant term $f_0$ is determined by:

$$f_0 = \int f(x) \prod_{i=1}^{m} [p_i(x_i) dx_i]$$
(19)

Therefore, the general form of k-order term of $f(x_1, x_2, ..., x_m)$ (a decomposition term depending on k input variables) is given by:

$$f_{i_1...i_m}(x_{i_1}, ..., x_{i_m}) = \int f(x) \prod_{j \neq i_1,...i_m} [p_j(x_j) dx_j] - \sum_{k=1}^{m-1} \sum_{j_1,...,j_k \in (i_1,...,i_m)} f_{j_1...j_k}(x_{j_1}, ...x_{j_k}) - f_0$$
(20)

A key assumption of Sobol's method is orthogonality, i.e., the terms of $f(x_1, x_2, ..., x_m)$ are uncorrelated with each other. As a result, the variance of $f(x_1, x_2, ..., x_m)$ can be determined by:

$$D = \sum_{i=1}^{m} D_i + \sum_{i_1=1}^{m} \sum_{i_2=i_1+1}^{m} D_{i_1 i_2} + ... + D_{1,...,m}$$
(21)

Sensitivity indices are then defined as:

$$S_{i_1,...,i_k} = \frac{D_{i_1,...,i_k}}{D}$$
(22)

And the summation of all the sensitivity indices equals 1:

$$\sum_{k=1}^{n} \sum_{i_1 < ... < i_k}^{n} S_{i_1,...,i_k} = 1$$
(23)

If k=1, then $S_{i_1\ldots i_k}$ is called main sensitivity index (MSI); if k ≥2, then $S_{i_1\ldots i_k}$ is called interaction sensitivity index (ISI). The total sensitivity index (TSI) is then defined as:

$$S_i^{tot} = S_i + \widehat{S}_{i,\sim i} = 1 - \widehat{S}_{\sim i} \qquad (24)$$

Where $\widehat{S}_{i,\sim i}$ is the summation of all the $S_{i_1\ldots i_k}$ that involve the index i and at least one index from (1,…, i-1, i+1, … m); $\widehat{S}_{\sim i}$ is the summation of all the $S_{i_1\ldots i_k}$ that do not involve any index i. $S_i^{tot}$ therefore represents the average variation in the outputs of the model that is contributable to the input variable i through its sole influences and interactions with other variables. Sobol's TSI requires heavy calculations of variance D and Di. To calculate D and Di, the marginal explained variance of output Y due to newly added X should be calculated recursively, following:

$$f_0 = E(Y) \qquad (25)$$

$$f_i(X_i) = E(Y|X_i) - f_0 \qquad (26)$$

$$f_{ij}(X_i, X_j) = E(Y|X_i, X_j) - f_0 - f_i - f_j \qquad (27)$$

Thus, when the user completed parameterizing $X_i$, $f_i$ can be calculated; when the user when the user completed parameterizing $X_j$, $f_{ij}$ can be calculated etc. The calculation process is repeated until $f_{1\ldots M}$ is calculated where M is the dimensionality of problem P. In this way, the calculation of Sobol's TSI is integrated with model definition and parameterization process.

**SUMMARY**

This paper presents a new method to compute VaR (value at risk) and corresponding variance based sensitivity analysis. VaR has a long history of being applied in stock price prediction and investment portfolio analysis. Traditional method is mainly analytical, has certain limitations.

The VaR simulation, on the other hand, provides more realistic analysis but is very slow which affects the applications. This study proposes a new VaR computation method based on a probabilistic simulation technique called Simulation As You Operate (SAYO). It is always helpful to know the most influential factors in an investment, and thus a sensitivity analysis method based on SAYO is also introduced to enhance investment analysis. Result indicates that the proposed new method is able to yield VaR and sensitivity analysis in real time.